\documentclass[aps, pre, twocolumn, epsfig ]{revtex4-2}

\usepackage{setspace}
\usepackage{times}

\usepackage{appendix}
\usepackage{bm}
\usepackage{graphicx}  
\usepackage{color}
\usepackage[dvipsnames]{xcolor}
\usepackage{natbib}
\usepackage{array}
\usepackage{graphics}
\usepackage{graphicx}
\usepackage{epsfig}
\usepackage{latexsym}
\usepackage{epstopdf} 
\usepackage{amsfonts}
\usepackage{graphicx}
\usepackage{grffile}
\usepackage{float}

\usepackage{array}
\usepackage{multirow}
\usepackage[english]{babel}
\usepackage[autostyle]{csquotes}
\usepackage[export]{adjustbox}
\usepackage{xcolor}
\usepackage{color}
\usepackage{amsmath}
\usepackage{mathtools}

\usepackage{hyphenat}
\let\oldref\ref
\renewcommand{\ref}[1]{(\oldref{#1})}
\usepackage[colorlinks=true,linkcolor=blue,urlcolor=blue,citecolor=blue]{hyperref}
\usepackage[normalem]{ulem}

\begin{document}
\title{Improved Desalination by Polymer Grafting}
\author{{Mamta Yadav*, Clifford E. Woodward** and Jan Forsman*}
\\  
{**}School of Chemistry
\\
University College,  University of New South Wales, ADFA
\\
Canberra ACT 2600, Australia   
\\
and
\\
{*}Theoretical Chemistry, Chemical Centre
\\
 P.O.Box 124, S-221 00 Lund, Sweden
\\
}
\begin{abstract}
  Freshwater scarcity demands desalination technologies that are efficient, scalable, and sustainable. Capacitive deionisation (CDI) is promising but remains
  limited by inefficient ion adsorption and poor charge utilisation. Here, we show that suitably chosen polyampholytic
  block copolymer grafting can substantially enhance CDI performance, via a combination of dipolar response and steric effects. 
  Using mean-field classical density functional theory and grand-canonical
  Monte Carlo simulations, we demonstrate that such polymer grafted electrodes enable strongly improved desalination performance, without altering
  the pore architecture. Even an electrode grafting by simple neutral polymers can generate an improvement, although a suitably designed
  block polymer architecture offers an additional performance gain.
  These results establish interfacial block copolymer grafting as a powerful route toward high-performance, membrane-free desalination.
\end{abstract}
\maketitle
{\em Introduction-} \label{IN}
\setcounter{equation}{0}
\renewcommand{\theequation}{1.\arabic{equation}}
Water is the foundation of life, yet access to safe drinking water is rapidly declining across the globe \cite{UNWater2024}. From remote
rural regions to densely populated urban centers, communities are struggling to secure reliable access to potable
water \cite{WHO_UNICEF2023, Rosa2022,MekonnenHoekstra2016}. Growing populations, industrial expansion, and agricultural
intensification are placing unprecedented pressure on limited freshwater resources \cite{Vrsmarty2010,WWAP2023}. Despite the
fact that nearly $70\%$ of the Earth is covered by water, most of it is saline and not suitable for direct human use. This contradiction
between abundance and scarcity represents one of the most pressing scientific and societal challenges of our time \cite{Shannon2008,USGSWater}. Traditional
water management systems are increasingly inadequate in the face of climate change and rising demand. Established desalination
technologies, including thermal distillation \cite{Prajapati2022} , reverse osmosis \cite{Greenlee2009}, and
electrodialysis \cite{AranaJuve2022}, have enabled large scale freshwater production. 
However, their high energy demand, operational costs, and environmental impacts limit long term sustainability, which motivates the development of
greener alternatives.  Developing sustainable technologies to convert seawater into potable water is therefore not just an
option, but a necessity for long term global resilience \cite{BENROUANE2025, Orfi2025}.

Capacitive deionisation (CDI) has emerged as an energy efficient and cost effective desalination strategy, offering a promising
route toward scalable and sustainable freshwater production \cite{Gude2016,Khalil2025}. By exploiting electric double layer
charging in porous electrodes, CDI enables the reversible electrosorption and release of ions under an applied
potential \cite{Murphy1967,Zou2008,Boon:11,Hartel:15}.

Conventional CDI electrodes based on activated carbon are fundamentally limited by low charge
efficiency. Considerable effort has therefore focused on improving electrode
materials \cite{Suss2015,Porada2013}. Nanostructured carbons such as graphene derivatives and carbon nanotubes provide
enhanced conductivity and an increased accessible surface area, while emerging materials including MXenes offer larger salt adsorption
capacities \cite{Das2014,Homaeigohar2017,Khan2023,Cheng2024}. Despite these advances, CDI performance remains constrained by nonselective ion desorption.
Membrane CDI (MCDI), partially addresses this issue by introducing ion exchange membranes that suppress
co-ion leakage and enhance salt removal efficiency \cite{Lee2006}. Similarly, flow-electrode CDI (FCDI) enables continuous
desalination by electrosorbing ions onto circulating carbon slurries, overcoming electrode saturation \cite{Yang2021}. However, FCDI
suffers from low charge efficiency and electrode degradation, while MCDI systems are often limited by weak electrode membrane
adhesion and increased contact resistance. Even recent designs incorporating ion-exchange polymers yield only marginal improvements
in desalination efficiency \cite{Lee2011}. These limitations highlight the need for fundamentally new strategies to control ion
adsorption at electrode interfaces \cite{Ahmad2024,Regenspurg2024}.

Block co-polymer based porous carbon fibers have recently shown improved CDI performance by optimising ion accessible surface area and
transport routes. However, these approaches rely primarily on bulk structural optimisation of electrode materials \cite{Liu2020,Kang2024}. In
contrast, polymer grafting provides a direct route to control interfacial chemistry and ion-surface interactions without altering the pore
architecture. A recent experimental study has shown that grafting conductive polymers onto activated carbon and  MnO$_2$ electrodes enhances
both desalination capacity and kinetics through pseudocapacitive ion storage \cite{Tan2020}. These results establish interfacial polymer
functionalisation as a promising route toward high-performance, membrane free capacitive deionisation. In this Letter, we demonstrate
that block copolymer grafted electrodes fundamentally reshape ion adsorption and
significantly enhance desalination performance. This finding is supported by desalination loops obtained
from classical density functional theory (cDFT), as well as by grand-canonical Monte Carlo simulations. Despite the coarse-grained nature of our
approach, the excellent agreement between these approaches highlights the robustness and predictive
power of the mean-field cDFT framework. Surprisingly, we have 
found that even neutral polymer grafts can deliver increased desalination efficiencies, although
a block charge polymer architecture leads to an additional gain.

{\textbf{\em Model and theory}}.
We settle with investigations of the negative electrode, and
propose a ``reversed'' treatment at the positive electrode.
We consider electrode pores, that are modelled as
slits, where planar and parallel walls of infinite size are separated a distance $H$.
The walls carry a uniform surface charge density $\sigma_s$, the
value of which is determined by the electroneutrality condition, for
an input value of the applied electrode potential.

Borrowing notation from earlier cDFT work \cite{Hartel:15}, we denote the electrode
volume by $V_e$ and the ``bucket'', to be desalinated,
volume by $V_b$. It is useful to define a ratio parameter $\xi = V_e/(V_e+V_b)$.
We shall work under $\Psi$-control, where $\Psi$ is the applied
electrode (Donnan) potential. We will compare two main cases, where the electrodes are either bare (reference)
or carry grafted polymers, that are overall electroneutral. For the main part of this work, the
grafted chains will have a block charge structure, where
the end monomer of the positive block (with one bonded neighbouring monomer) is
grafted to the electrode.  We might, for instance, write \newline
$|+ + + + - - - -$  \newline
as a pictorial
illustration of grafted 8-mers (4+4), where ``$|$'' describes the planar electrode surface
to which the ``left-most'' positive monomer is grafted.
We will, however, also explore the option of grafted polymers where
all monomers are electroneutral. 
In all cases, the fixed bond length (but rotationally flexible) between
connected monomers is set to $b=6${\AA}. Moreover, in all cases
of the main paper,  all simple ions and monomers have a hard-sphere (HS) diameter of $d=4${\AA}.
In {\em Appendix B}, we will also explore the option of vanishing hard cores, for all species.
The densities of positive and negative monomers are denoted $n_+$ and $n_-$.

In the slit geometry, where the planar charged surfaces are separated by $H = h+d$
we employ the following grand
potential per unit area $\omega$, for the case of grafted
$r-mer$ chains, composed of charged blocks:
\begin{widetext}
\begin{equation}
  \begin{split}
    \beta \omega(T,\phi_g,\mu_c,\mu_a,\Psi;[N({\bf Z}), n_c(z), n_a(z)])  =  \quad \\        
    \int N({\bf Z}) \left(\ln [N({\bf Z})] - 1 \right) d{\bf Z} + \beta \int N({\bf Z}) V_b({\bf Z}) d{\bf Z}
  +\int n_c(z)\left(\ln[n_c(z)]-1\right) + n_a(z)\left(\ln[n_a(z)]-1\right) dz + \\
  +\beta \int {\cal F}_{HS}^{(ex)}(n_+,n_-,n_c,n_a) dz+\frac{\beta e}{2}  \int \psi(z) q(z) dz  
    -\beta \int(\mu_c n_c(z) + \mu_a n_a(z))dz + \beta \int n_+^e(z) \phi_g dz
\end{split}
\end{equation}
\end{widetext}
where $e$ is the elementary charge, $z$ denotes the coordinate perpendicular to the surfaces,
${\bf Z} = z_1...z_r$, $N({\bf Z})$ is polymer density distribution, and $\psi(z)$ is the local
mean-field potential at $z$, {\em including} the
applied electrode potential $\Psi$. The grafting potential $\phi_g$ only acts on positive
end monomers, with a density $n_+^e(z)$.
In all cases reported in this work, we have used a fixed grafting
density, such that $\int_0^{H/2} n_+^e(z) dz = 0.02/${\AA}$^2$. 
The anion and cation densities are
denoted $n_a$ and $n_c$, and their chemical potentials are $\mu_a$ and $\mu_c$ (note that $\mu_a=\mu_c$ since
we adopt an RPM). The net charge distribution at $z$ is denoted $q(z)$, i.e.
$q(z) = e(n_c(z)+n_+(z)-n_a(z)-n_-(z))+\sigma_s(\delta(z)+\delta(H-z))$.
Excluded volume effects are managed by ${\cal F}_{HS}^{(ex)}(n_+,n_-,n_c,n_a)$, where we have
utilised the Generalised Flory-Dimer approximations \cite{Forsman:2004b}.
The centres of mass of monomers and simple ions are confined
to the region $d/2 < z < H-d/2$.
All calculations were performed at a temperature $T=298$K, representing the
aqueous solvent implicitly, via the dielectric constant, $\epsilon_r=78.3$.
The equilibrium grand potential per unit area, $\omega_{eq}(h)$, was obtained by numerical minimisation by
Picard iterations, using methods for polymer-cDFT that are well established \cite{Woodward:1991,Xie:2016}.

The desalination loop starts at $\Psi = 0$.
With $<n_c>$ and $<n_a>$ representing the average (free) cation and
anion concentrations within $V_e$ we have, at $\Psi = 0$ and under full
grand canonical conditions:
\begin{equation}
  n_{tot} = 2n_s(1-\xi) + \xi(<n_c>+<n_a>)
\end{equation}
where $n_s$ is the salt concentration in the bucket (initially at 0.6M)
and $n_{tot}$ represents the overall density of simple ions, as averaged across
the bucket and electrode volumes.
Thus, $n_{tot}$ is determined at this initial state, for a given
set of system parameters, such as
the presence of grafted polymers (or not), surface separation($H$), $\xi$-value etc. 

From this starting state, we turn on a (negative) Donnan potential
and trace two different paths:
\begin{itemize}
\item A full grand canonical path where there is complete chemical
  equilibrium with a bulk at a fixed concentration of 600mM.
\item A desalination path, where we gradually drain
  the ``bulk'', i.e. the chemical potential is allowed to
  vary, and the appropriate value has to be found by an
  iterative procedure, where
\begin{equation}
     n_{s}(\mu_c,\Psi) = \frac{1}{2(1-\xi)}(n_{tot}-\xi(<n_c(\mu_c^{eff})>+<n_a(\mu_a^{eff})>))
  \label{eq:iter}
\end{equation}
where $\mu_c^{eff} = \mu_c-e\Psi$ and $\mu_a^{eff} = \mu_a+e\Psi$
with $e$ denoting the elementary charge and $\mu_c=\mu_a$ representing the ion chemical
potentials in the bulk.
This is repeated until the bucket concentration, $n_s$, is commensurate with the
bulk ion chemical potential. The desalination process ends when $n_s$ has reached the target
concentration, which in this work is 1.2e-5/{\AA}$^3$, i.e. about 20mM.
\end{itemize}
The full cycle has been nicely illustrated via a cartoon, in Figure 5 (e) of
ref. \cite{Hartel:15}

\begin{figure}[h!]
\begin{center}
       \includegraphics[scale=0.34]{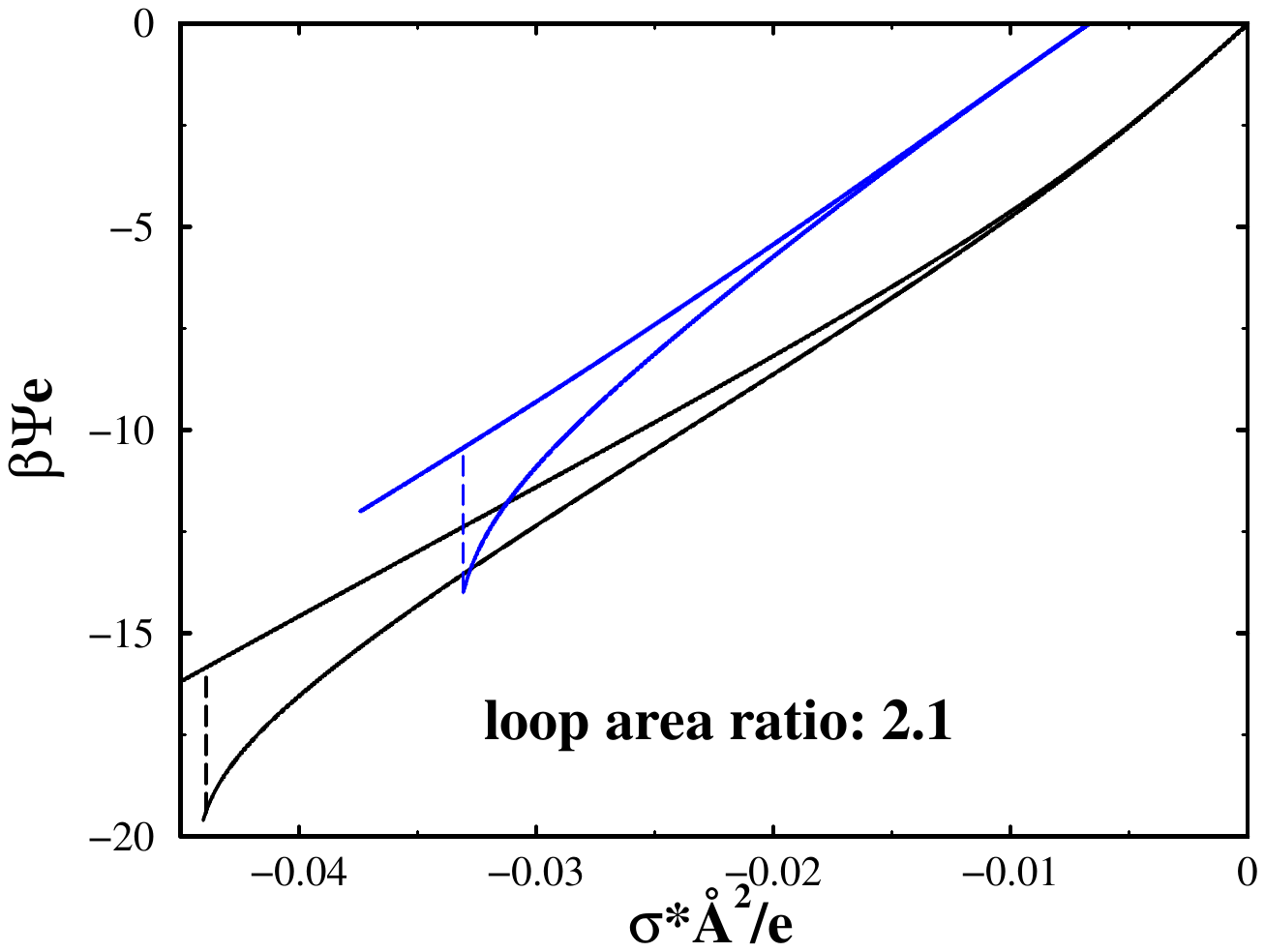}  
       \caption{\footnotesize{cDFT predictions of desalination curves, in the
           presence, and absence, of grafted 8-mer block chains (4+4).
           The surfaces are separated by $H=80${\AA}, and $\xi=0.65$, which means that
           the model system is the same as in Figure \ref{fig:MC} of the Appendix.
           The dashed lines denote the rinsing step, where
           the concentration in the salt-drained bucket
           has dropped from the initial value of 600mM to $n_s = 1.2*10^{-5}/${\AA}$^3$ (20mM).
           The ratio between the enclosed (loop) areas is about 2.1.           
}}
\label{fig:cDFT}
\end{center}
\end{figure}

\textbf{{\em An example of a desalination loop}.}
Let us start with a specific model electrode, where $H=80${\AA}. 
cDFT predictions for the desalination loop in the
presence, and absence, of electrode-grafted polymers, are shown in Figure \ref{fig:cDFT}.
We see that, not only is
the area in the reference case more than twice as large
as with grafted electrodes, but we also need to apply a stronger
potential to reach our target.

The above description corresponds to a mean-field treatment of electrostatics. This is
motivated by simplicity, and the fact that all charges are monovalent. We have established
additional support by direct comparisons of desalination loops, as predicted
by cDFT and grand canonical Metropolis Monte Carlo (MC) simulations.
Results from the latter approach are reported in {\em Appendix A}
and we find the agreement satisfactory, especially when considering the coarseness
of our model and the largely qualitative nature of our conclusions.
We will from now on proceed with predictions by mean-field cDFT.
 
{ \textbf{\em Simple ion density profiles}.}
Let us continue by examining some density profiles at high
and low applied potentials, comparing data for
bare and polymer-covered electrodes.
\begin{figure}[h!]
    \includegraphics[scale=0.45]{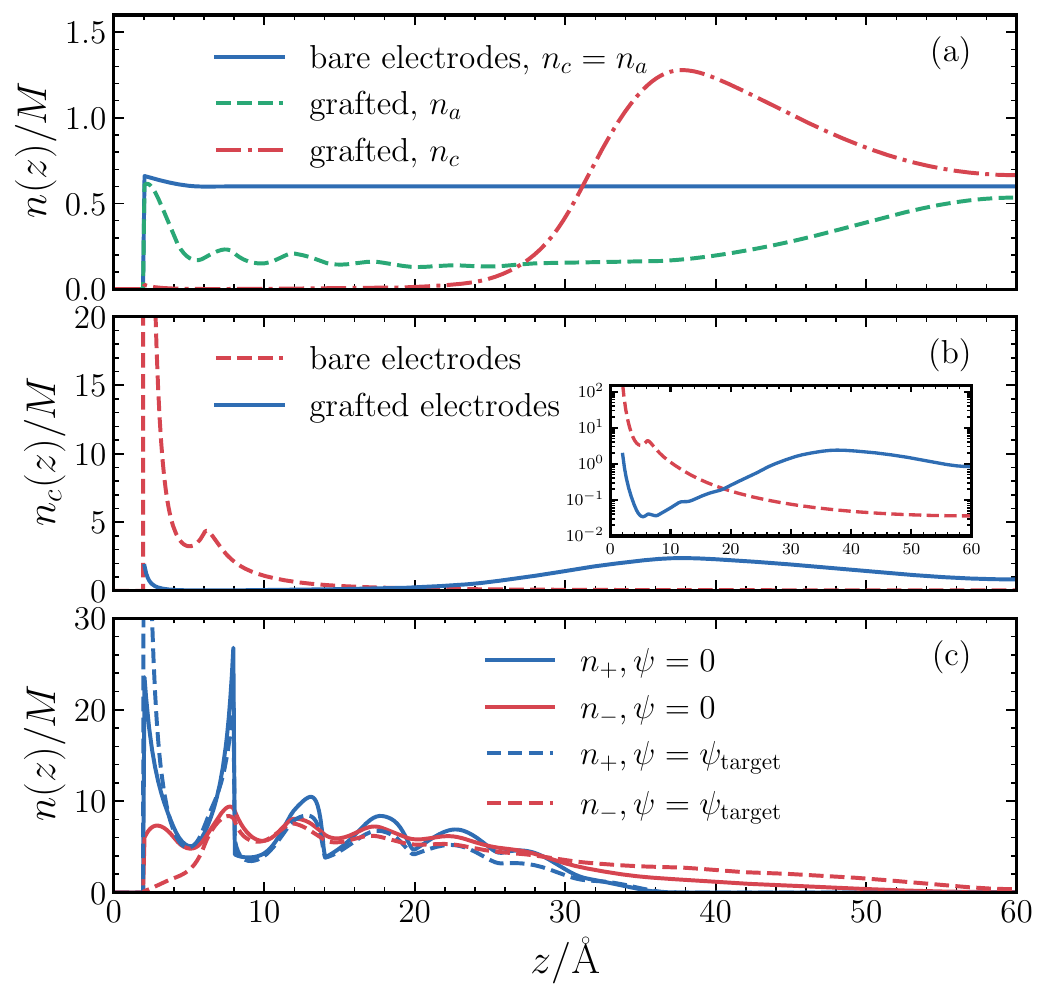}    
 \caption{\footnotesize{Ion distribution profiles, at $H=120${\AA} and $\xi=0.8$, with bare and polymer grafted electrodes. In the latter case, the
   chains are 12-mers. The applied potential is zero in graph (a), and at the value ($\Psi_{target}$) for which the bucket salt
   concentration has dropped to the target value, in graph (b).This means that the absolute value of the Donnan
   potential is stronger for bare electrodes than those for polymer-grafted (in graph(b)). Graph (c) displays
 monomer density profiles.}}
\label{fig:dens}
\end{figure}
In Figure \ref{fig:dens} (a) we see how the polymers effectively generate a
double-layer even at zero applied potential. Moreover, there
is a relative accumulation of cations under these conditions, with anions
being depleted from the slit region. While the cation concentration is
vanishingly small near the walls (due to excluded volume and a
layer of positive monomer charge), there is a region outside the
polymer layer, where the anions accumulate, on account of their
affinity to the negative outer part of the polymer brush.
A direct consequence of these distributions at zero applied
potential, is a negative surface charge.

In Figure \ref{fig:dens}(b), we display
cation density profiles at the ``turning point'' of the desalination loop (just prior to the rinsing step, denoted
by dashed vertical lines in Figure \ref{fig:cDFT}), i.e. where
the bucket salt concentration has dropped to the target value 20mM, for
bare and polymer-covered electrodes (the anions have a very low concentration in
both cases). Given that it is costly to generate density gradients, it
is perhaps not too surprising that the charging free energy cost
with polymer-grafted electrodes is, as we shall see, lower than for the bare ones.

{ \textbf{\em Comparisons of maximum performance}.}
Before we proceed, it is useful to derive a
systematic and unambiguous measure of ``performance''.
We can monitor the net uptake of ions within the slit as:
\begin{equation}
  <N_c>_f-<N_c>_s + <N_a>_f-<N_a>_s
\end{equation}
where indices ``$f$'' and ``$s$'' stands for ``final'' and ``start''.
We can measure (say) $<N_c>_f$ as $(\int n_c(final) (z) dz)A$, where $A$ is the area. A reasonable
measure of ``performance'', $\gamma$, is how many ions that can be extracted from the bucket,
for a given energy input, i.e.: 
\begin{equation}
  \gamma = \frac{(<N_c>_f-<N_c>_s + <N_a>_f-<N_a>_s)}{\Delta w}
  \label{eq:gamma}
\end{equation}
where $\Delta w$ is the net work of a cycle, i.e. the area between
the two curves in the desalination process.

\begin{figure}[h!] 
 \includegraphics[scale=0.45]{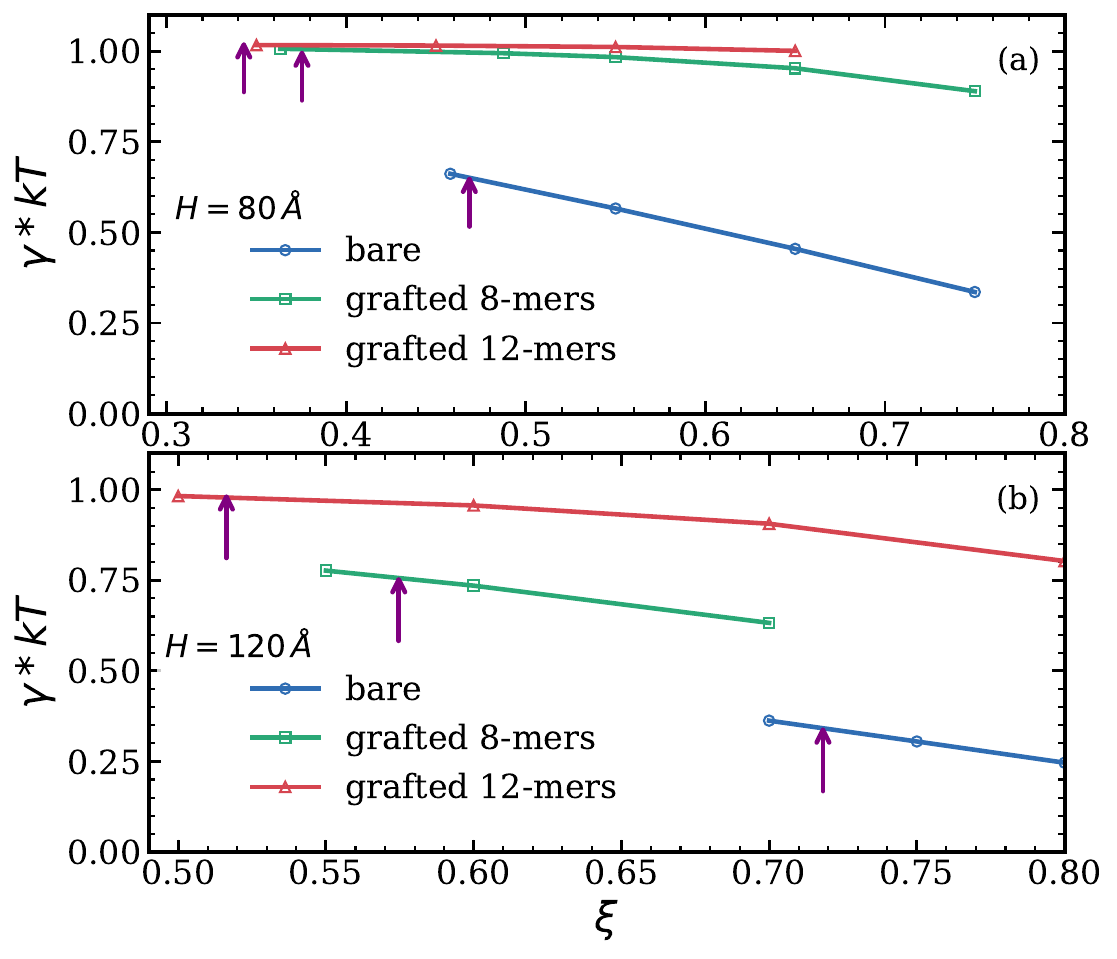}    
  
  \caption{\footnotesize{The dependence of $\gamma$ on $\xi$ at various separations, for  
    bare and grafted surfaces. The arrows indicate the electrochemically
    limited maximum performance, $\gamma_{max}$, in each case. 
      }}
        \label{fig:gammavsxi}          
\end{figure}

The performance will obviously depend on the ratio $\xi$, but also
on the separation. Examples are provided in Figure \ref{fig:gammavsxi}. 
The arrows in the graphs mark the points where $\Psi = -|\Psi|_{max} = -0.615$V, i.e.
the maximum attainable performance at the given separation, as limited
by the electrochemical window in aqueous environments (1.23V between the electrodes). 
It is clear that the maximum performance benefits from the
presence of the grafted block polymers.
There are probably several reasons
for this. We have already mentioned the dramatic drop of
density gradients found when the polymer layers are in place. 
Another reason for the improved performance is the dipolar response afforded
by the block polymer layer. It should be noted that if we
assign a higher dielectric constant to our implicit solvent, we also
arrive at an increased performance (not shown). Finally, there is
a steric aspect that we shall analyse in more detail below.
In passing, we mention that we have made tests with a
reduced grafting density, resulting in a small drop of the desalination performance (not shown).

One interesting observation from Figure \ref{fig:gammavsxi} is that the
variation of $\gamma$ with $\xi$ tends to be more flat
with polymer-grafted electrodes. This may be useful in scenarios where
a higher $\xi$, i.e. weaker $|\Psi|_{max}$, is desirable, perhaps
for kinetic reasons, or simply because one prefers some safety margin
for possible onset of redox reactions.
\begin{figure}[h!]
\begin{center}
       \includegraphics[scale=0.34]{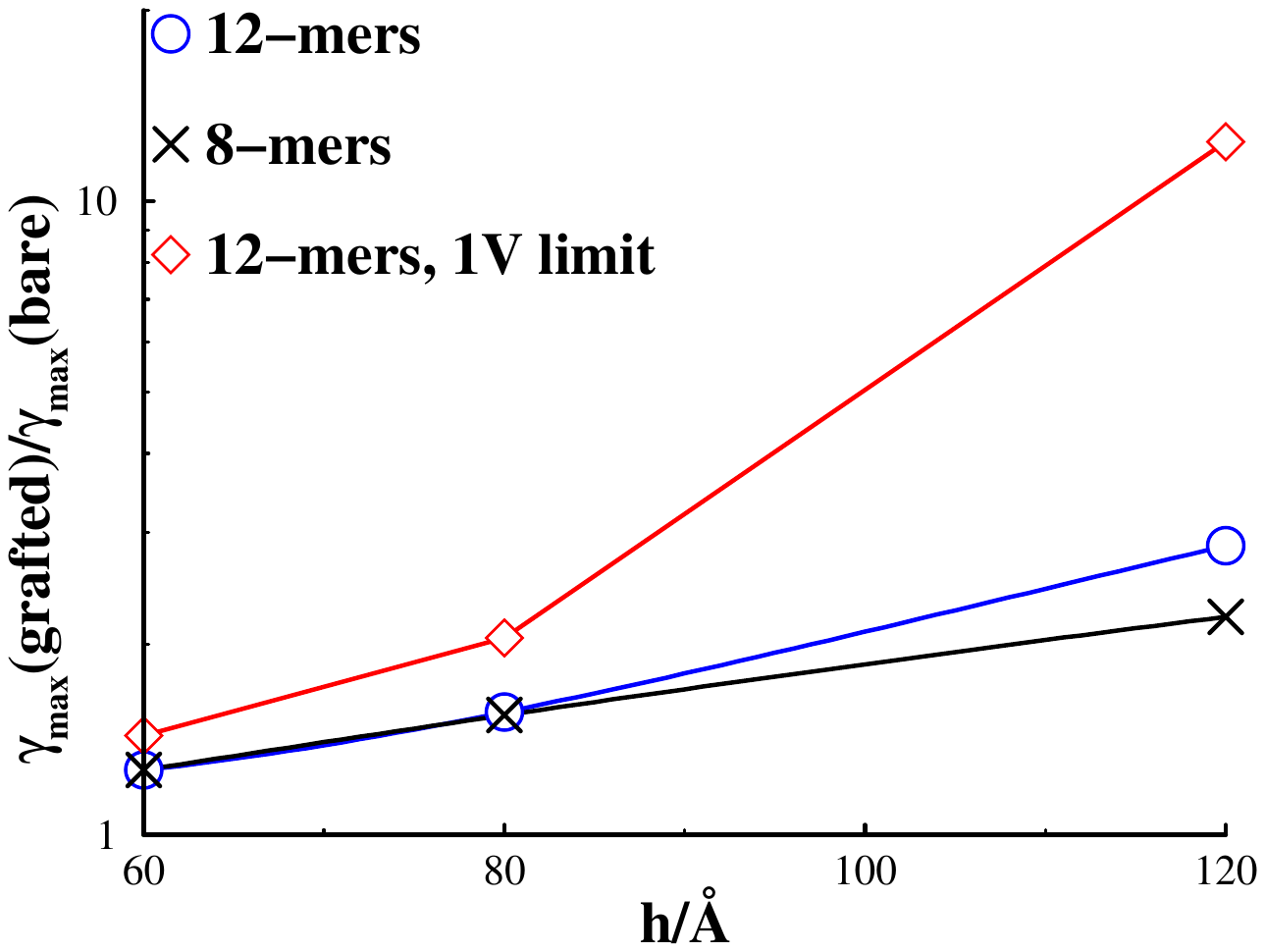}  
       \caption{\footnotesize{The ratio between $\gamma_{max}$, as obtained with
           block-polymer grafted and 
           bare electrodes.
           The solid curves are a guide to the eye.           
       }}
\label{fig:gammamax}
\end{center}
\end{figure}
Reducing  $|\Psi|_{max}$ can have a rather dramatic
effect, as is illustrated in Figure \ref{fig:gammamax}, where we plot
the ratio between the maximum performances with and without
polymers, at various separations.
With a ``conservative'' limit of 1V (between the electrodes), our
cDFT predicts an improvement to the maximum performance by more than
an order of magnitude(!), at a separation of 120{\AA}. 

{\textbf{\em Polymers composed of neutral HS monomers}.}
As stated above, there also seem to be steric mechanisms
that promote desalination performance. The presence of
dense polymer brushes can, at least to some degree, be
viewed as a drop of the ``effective'' separation.
From Figure \ref{fig:gammavsxi} it is obvious that
$\gamma_{max}$ increases as the separation drops, especially for
bare electrodes.
\begin{figure}[h!]
\begin{center}
       \includegraphics[scale=0.34]{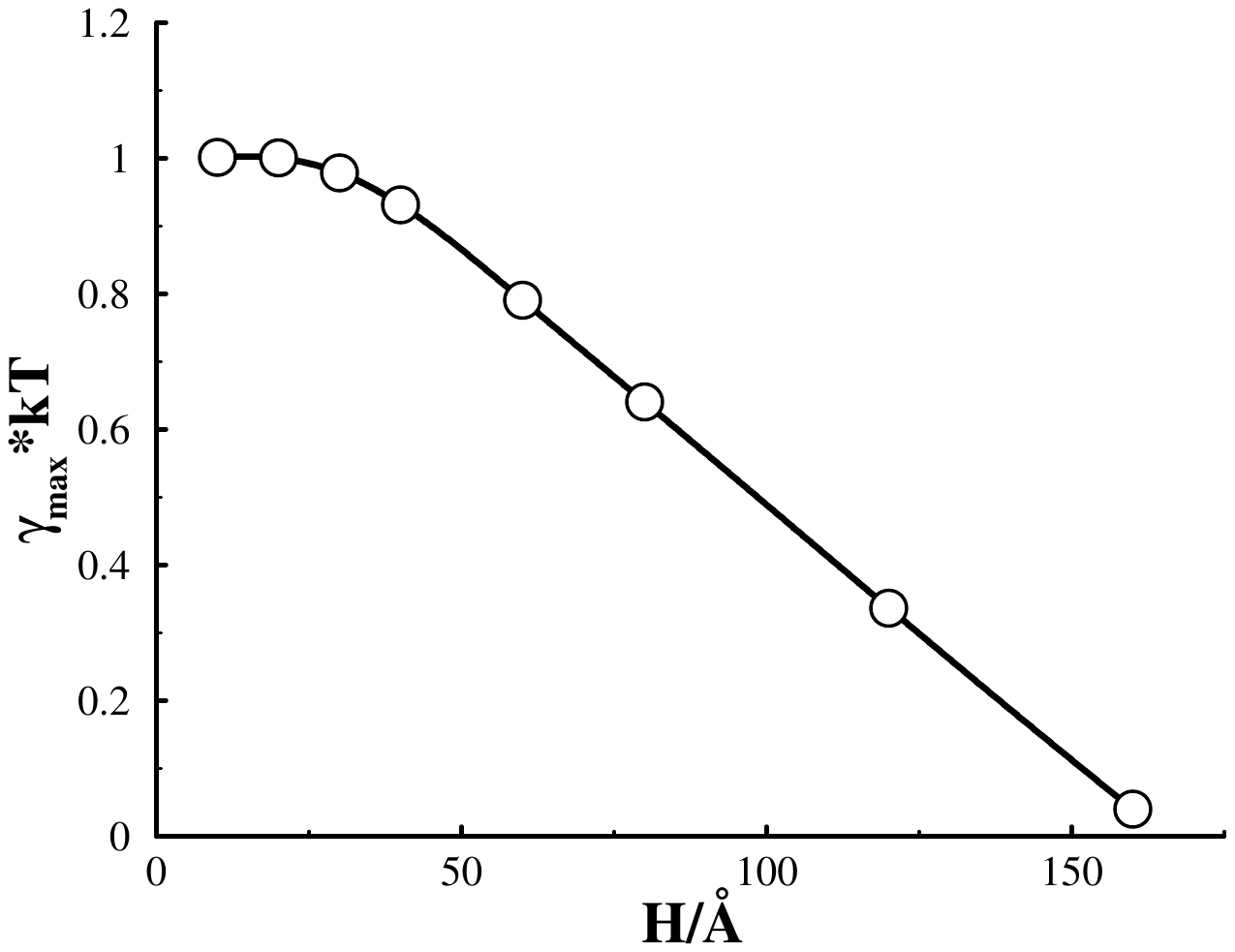}  
       \caption{\footnotesize{The variation of $\gamma_{max}$
           with separation, for bare electrodes. 
       }}
\label{fig:HSbare}
\end{center}
\end{figure}
This is quantified in Figure \ref{fig:HSbare}, where we note
a plateau at very small separations, with a monotonic
decline at larger slit widths. 
This suggests that it might be possible
to ``mimic'' a more narrow pore, with an increased performance, simply
by grafting inert polymers, the monomers of which merely exclude volume.
Indeed, the results presented in Figure \ref{fig:HS} corroborate this
conjecture.

\begin{figure}[h!]
\begin{center}
       \includegraphics[scale=0.34]{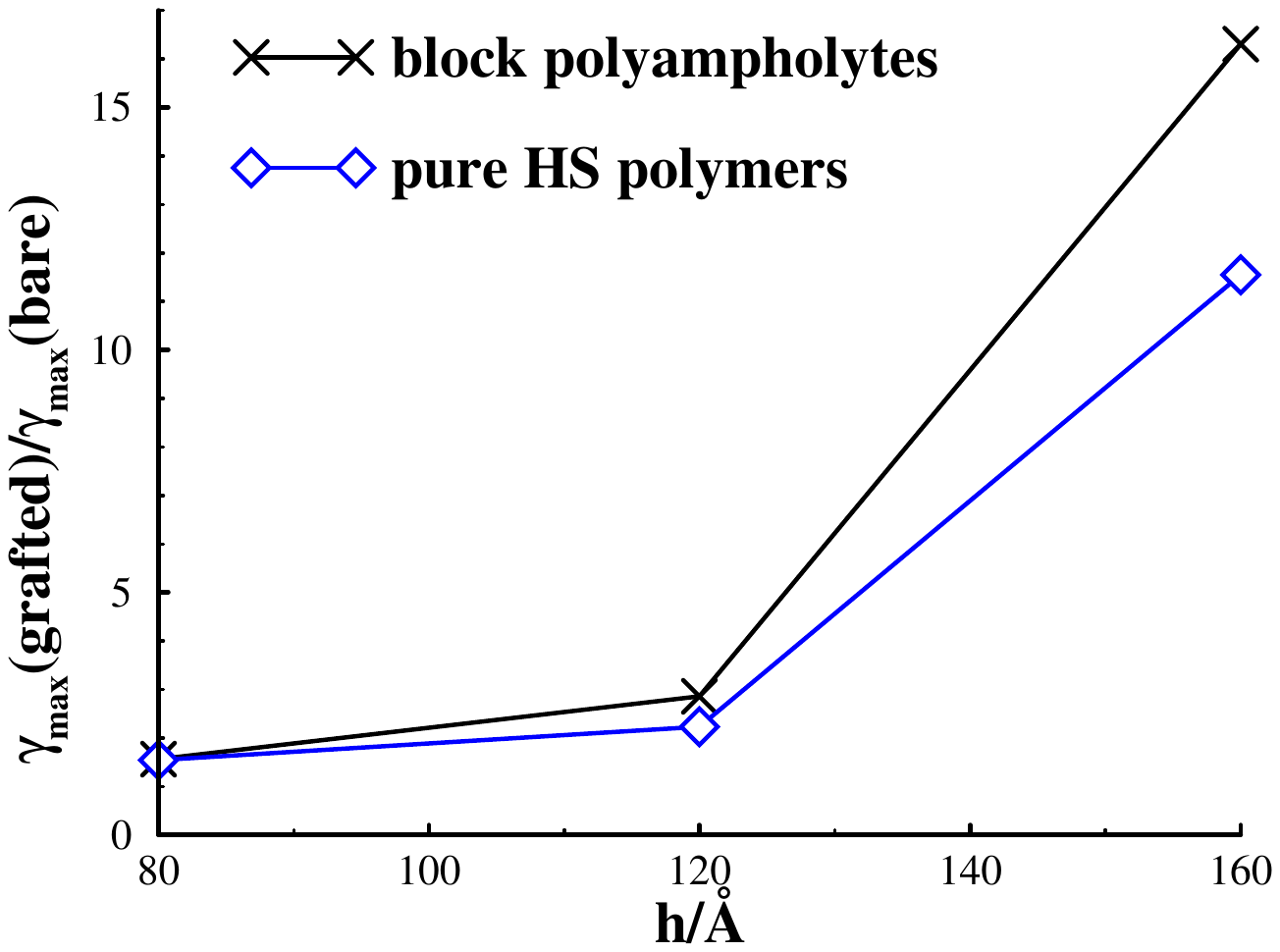}  
       \caption{\footnotesize{The ratio between $\gamma_{max}$, as obtained with
           polymer grafted (12-mers) and bare electrodes. Black crosses
           are results obtained with our ``standard''
           grafted polyampholytes (net neutral), with a block architecture.
           Blue diamonds, however, are results from calculations
           with grafted polymers in which {\em all monomers} are
           neutral (pure HS monomers).
       }}
\label{fig:HS}
\end{center}
\end{figure}

The results for 12-mer chains, composed of pure HS monomers are
virtually identical to those for 8-mer polyampholytes (compare
with results in Figure \ref{fig:gammamax}). At the largest separation reported, we see that block-like 12-mers do
provide an extra ``boost'' to the performance.
Still, if grafting by simple neutral chains
is cheaper or easier to achieve in practice, then this is certainly
a viable option, well worth exploring.
If one has porous electrodes with a rather wide pore size distribution,
it might be possible to establish a grafting procedure where the
polymers (pure HS or block polyampholytes) enter and form grafts within wide pores, but
are excluded from narrow pores. This exclusion is then, however, not
necessarily a problem since the desalination performance is high under those
conditions, even with bare electrode surfaces.

{\textbf{\em Acknowledgment}.} J.F. acknowledges financial support by the Swedish Research Council.	

\bibliographystyle{apsrev4-2}
\bibliography{draft}
\section*{End matter} 

{ \textbf{\em Appendix A: simulations}.}

\begin{figure}[h!]
\begin{center}
       \includegraphics[scale=0.34]{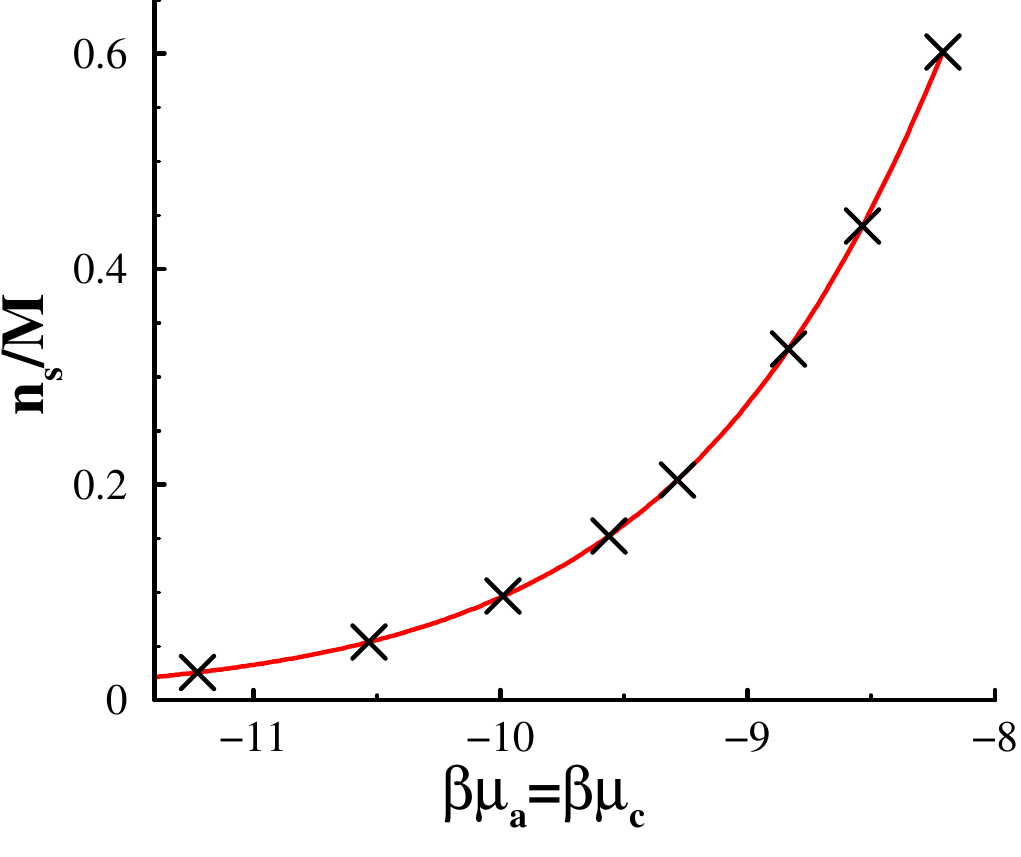}  
       \caption{\footnotesize{The dependence of the bucket
           salt concentration, $n_s$, on the single
           ion chemical potential, $\mu_c$ (or ($\mu_a$)).
           Symbols indicate simulated data points, and the red
           line marks a cubic spline. 
}}
\label{fig:bulkchemp}
\end{center}
\end{figure}

Here, we first need to numerically establish how the salt chemical
potential varies with salt concentration in a bulk solution. Since
we use Ewald GCMC methods, it is convenient to tabulate
how the bulk salt concentration, $n_s$, varies with the single ion (input) chemical
potential ($\mu_c$ or $\mu_a$, where $\mu_c=\mu_a$). 
The results are summarised in Figure \ref{fig:bulkchemp}.

Given that the dependence shown in Figure \ref{fig:bulkchemp} was established, we
proceeded with electrode pore simulations, with the pore walls (as in cDFT) being described
as flat and infinitely large walls. Specifically, we adopted a newly developed method
using single-ion insertions and deletions (GCMC), combined  with an applied electrode (Donnan) potential that
regulates the ion accumulation, and therefore the surface charge density. 
The conducting walls are treated using a 3D Ewald formulation of the Coulomb interactions 
to deal with the infinite number of image interactions that occur due to multiple reflections across the
electrode surfaces\cite{Stenberg:20}.   For the symmetric electrolyte model that we consider in this work, the individual
cation and anion chemical potentials are equal in the bulk solution ($\mu_a=\mu_c$). At each step of the desalination process, we
then need to reach consistency between the bucket salt concentration, $n_s$, as established by eq.  {\ref{eq:iter}, and the chosen ion chemical potential.  

In Figure \ref{fig:iter}, we illustrate an iterative search
for the appropriate chemical potential, and thus 
step (ii). In this case, a slight extrapolation was required
to reach the bulk ``master curve'', from which the bucket salt concentration
is obtained. It would obviously have been preferable with an interpolation, but
the extent of extrapolation is quite small, and thus not a major problem in
this case.
\begin{figure}[h!]
\begin{center}
       \includegraphics[scale=0.34]{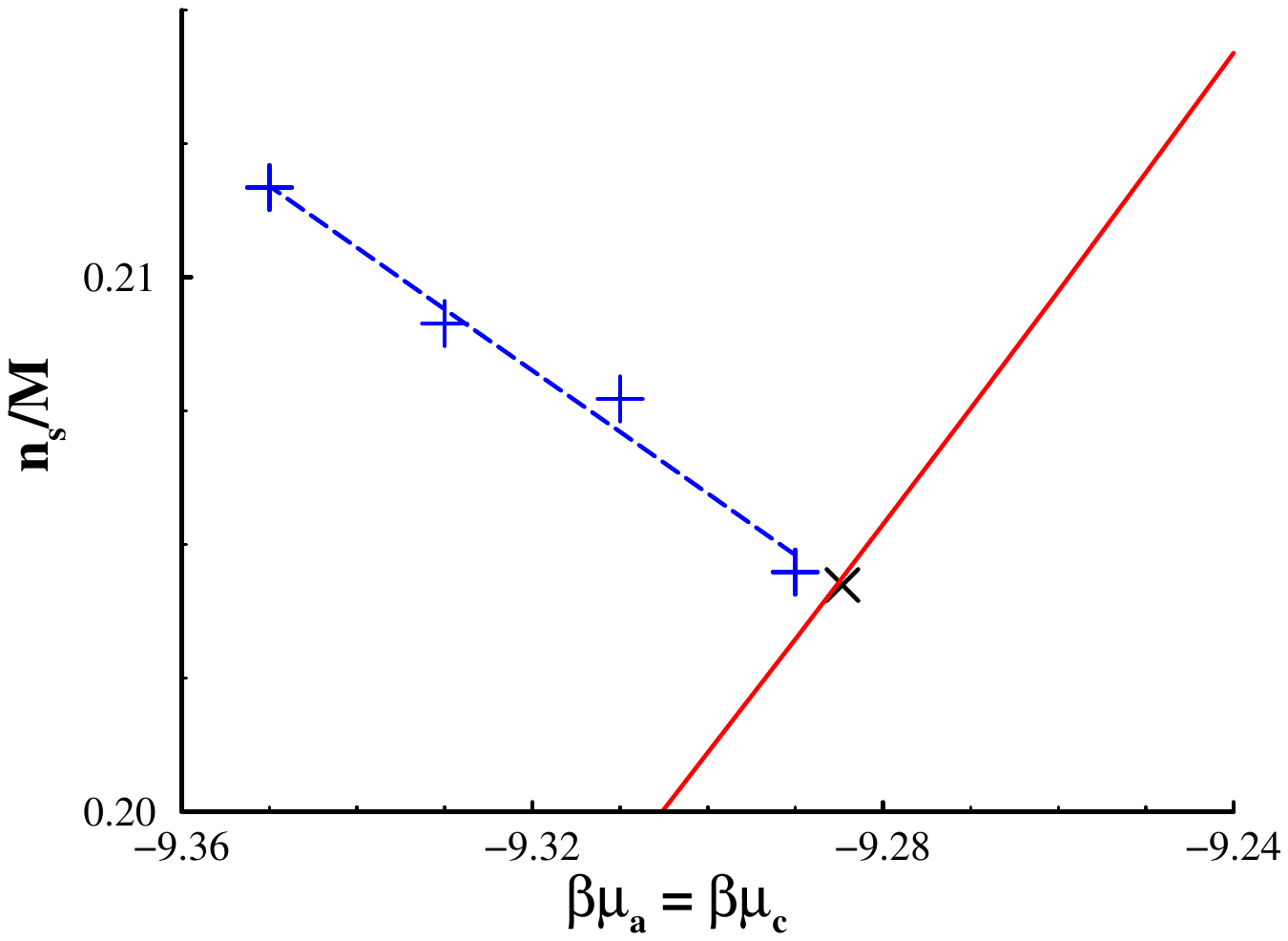}  
       \caption{\footnotesize{Simulated search for the correct
           chemical potential during desalination. In the displayed
           case (with grafted 8-mers), the Donnan potential is
           $\beta\Psi e = 8.0$. Blue plus signs denote simulated
           values of $n_s$, using average slit densities, and eq.(\ref{eq:iter}).
           The black cross symbol indicate the (extrapolated) target
           chemical potential.
}}
\label{fig:iter}
\end{center}
\end{figure}
\begin{figure}[h!]
\begin{center}
       \includegraphics[scale=0.34]{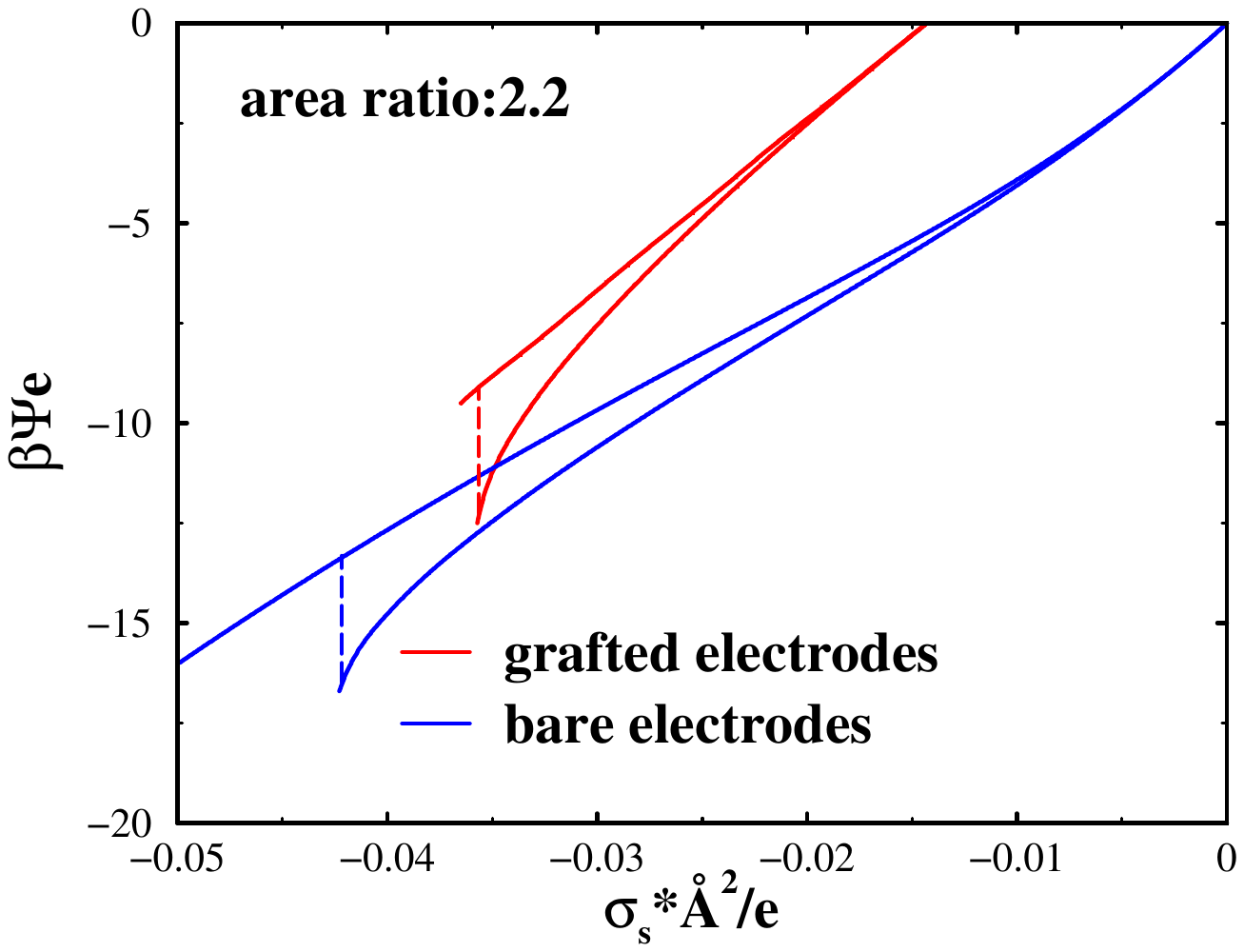}  
       \caption{\footnotesize{Simulated desalination curves, in the
           presence, and absence, of grafted 8-mer block chains.
           The surfaces are separated by $H=80${\AA}, and $\xi=0.65$.
           The dashed lines denote the rinsing step (subsequent to bucket water
           harvesting), where
           the concentration in the salt-drained bucket
           has dropped to target value.
           These results can be directly and quantitatively
           compared with those presented in Figure \ref{fig:cDFT}, where
           cDFT predictions for the same system are displayed. 
           The ratio between the enclosed (loop) areas
           is about 2.2.
}}
\label{fig:MC}
\end{center}
\end{figure}
We proceed in this manner, with a range of Donnan potentials, until
we reach our target, i.e. a Donnan potential strong enough to generate
$n_s = 1.2*10^{-5}/${\AA}$^3$, i.e. a bucket salt concentration of about 20mM. By doing this in the presence, and absence, of
grafted 8-mers, we can estimate predicted effects from grafting. An example is shown in Figure \ref{fig:MC}.

We are quite happy with the overall agreement with
the corresponding cDFT predictions, presented in Figure \ref{fig:cDFT},
even though it is far from perfect.

{\textbf{\em Appendix B: systems with vanishing excluded volume}.}
Here we make a brief digression to demonstrate that grafting by net neutral
polyampholytes with a block charge structure can improve
desalination performance also in the complete absence of
monomer and ion hard cores, i.e. there is no excluded
volume, except by the walls.
\begin{figure}[h!]
\begin{center}
       \includegraphics[scale=0.34]{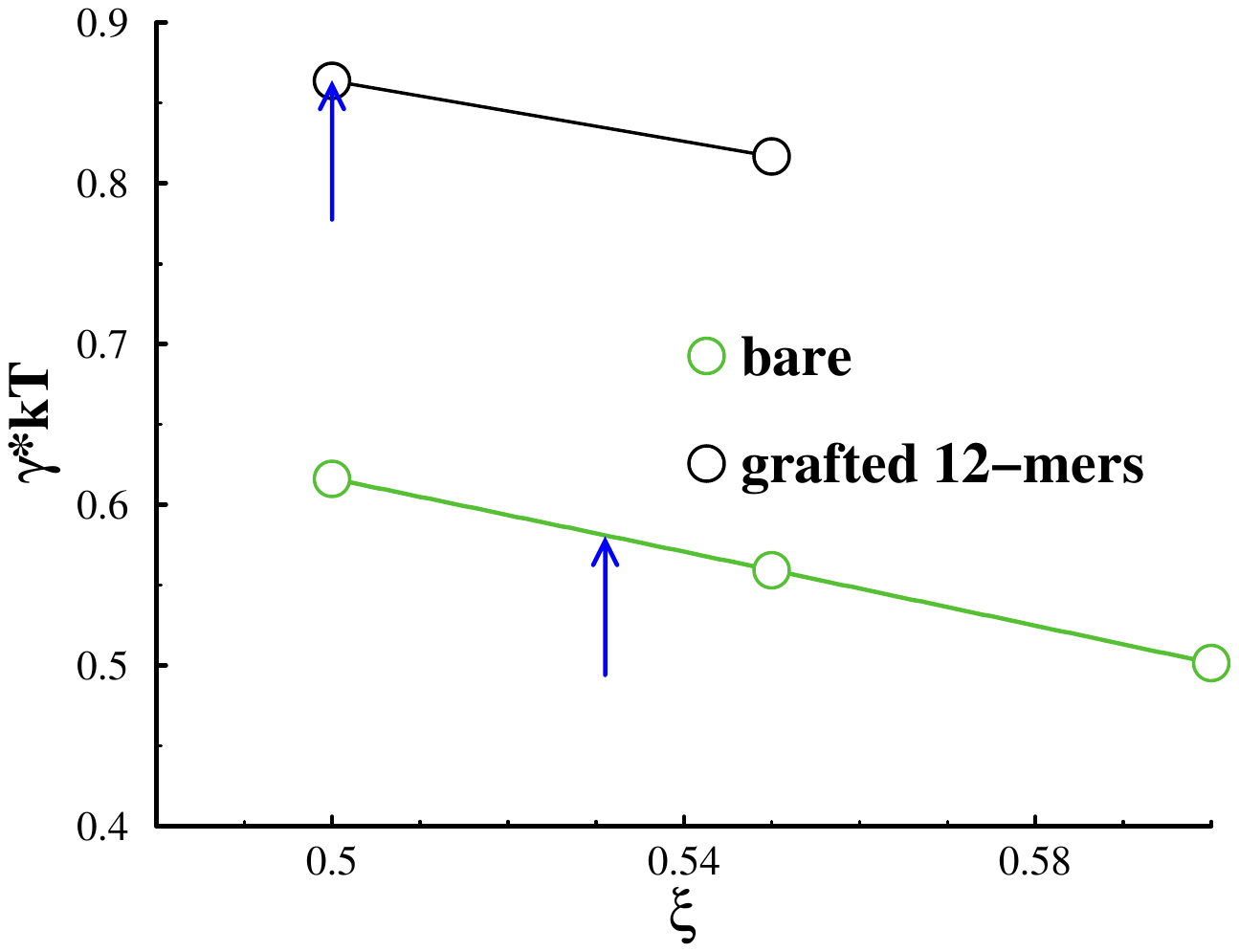}  
       \caption{\footnotesize{The dependence of $\gamma$ on $\xi$ at $H=120${\AA},
           for systems in which the particles (monomers and simple ions) are
           point-like, i.e. there is no hard-core excluded volume (save the walls).
       }}
\label{fig:ideal}
\end{center}
\end{figure}
In Figure \ref{fig:ideal} we see that, at a separation of
$120${\AA}, $\gamma_{max}$ increases by about 50 {\%} when the
electrodes are grafted. This clearly shows that while excluded
volume effects are important to the overall CDI performance, the
polar response offered by block chains leads to
an additional improvement.

\end{document}